\documentclass[onecolumn]{aa}  
\usepackage{amsmath}
\usepackage{multicol}

\usepackage{natbib,twoopt}
\usepackage[draft]{hyperref} 
\bibpunct{(}{)}{;}{a}{}{,}             
\makeatletter
  \newcommandtwoopt{\citeads}[3][][]{\href{http://adsabs.harvard.edu/abs/#3}%
    {\def\hyper@linkstart##1##2{}%
     \let\hyper@linkend\@empty\citealp[#1][#2]{#3}}}
  \newcommandtwoopt{\citepads}[3][][]{\href{http://adsabs.harvard.edu/abs/#3}%
    {\def\hyper@linkstart##1##2{}%
     \let\hyper@linkend\@empty\citep[#1][#2]{#3}}}
  \newcommandtwoopt{\citetads}[3][][]{\href{http://adsabs.harvard.edu/abs/#3}%
    {\def\hyper@linkstart##1##2{}%
     \let\hyper@linkend\@empty\citet[#1][#2]{#3}}}
  \newcommandtwoopt{\citeyearads}[3][][]%
    {\href{http://adsabs.harvard.edu/abs/#3}
    {\def\hyper@linkstart##1##2{}%
     \let\hyper@linkend\@empty\citeyear[#1][#2]{#3}}}
\makeatother

\usepackage{graphicx}
\usepackage{txfonts}
\usepackage{gensymb}
\usepackage{multirow}

\newcommand{\kms}{km$\,$s$^{-1}$ }
\newcommand{\nnh}{N$_2$H$^+$ }
\newcommand{\nqn}{N$^{15}$NH$^+$ }
\newcommand{\qnn}{$^{15}$NNH$^+$ }

\newcommand{\ratio}{$^{14}$N/$^{15}$N }

\newcommand{\isot}[2]{\ensuremath{^{#1}\mathrm{#2}}}
\newcommand{\ee}{\ensuremath{\mathrm{e}}}
\newcommand{\dd}{\ensuremath{\mathrm{d}}}
\newcommand{\wsxj}[6]{\ensuremath{\left\{\begin{matrix} #1 & #2 & #3 \\ #4 & #5 & #6 \end{matrix}\right\}}}

\begin{document}

   \title{\ratio ratio measurements in prestellar cores with \nnh: new evidence of $^{15}$N-antifractionation\thanks{This work is based on observations carried out with the IRAM 30 m Telescope. IRAM is supported by INSU/CNRS (France), MPG (Germany) and IGN (Spain).}}


   \author{E. Redaelli
          \inst{1}
          \and
         L. Bizzocchi\inst{1} \and
         P. Caselli \inst{1} \and
         J. Harju \inst{1,2} \and
         A. Chac\'on-Tanarro \inst{1} \and
           E. Leonardo \inst{3} \and
           L. Dore \inst{4}}
           
  \institute{Centre for Astrochemical Studies, Max-Planck-Institut f\"ur extraterrestrische
              Physik, Gie\ss enbachstra\ss e 1, D-85749 Garching bei M\"unchen (Germany) \\
              \email{[eredaelli,bizzocchi,caselli,harju,achacon]@mpe.mpg.de}
              \and
              Department of Physics, PO Box 64, 00014 University of Helsinki, Finland \and
              Istituto de Astrof\'isica e Ci\^encias do Espa\c co,
              Universidade de Lisboa, OAL, Tapada da Ajuda, PT1349-018 Lisboa (Portugal)
              \email{eleonardo@oal.ul.pt}
              \and
              Dipartimento di Chimica ``G.~Ciamician'',  Universit\`a di Bologna,
              via F.~Selmi 2, I-40126 Bologna (Italy).
              \email{luca.dore@unibo.it}
             }

     \titlerunning{\isot{14}{N}/\isot{15}{N} isotopic ratio in prestellar cores}
     
   \authorrunning{Redaelli et al.}
   \date{Received ****; accepted *****}

 
  \abstract
   {
 The $^{15}$N fractionation has been observed to show large variations among astrophysical sources, depending both on the type of target and on the molecular tracer used. These variations cannot be reproduced by the current chemical models. }
   {Until now, the \ratio ratio in \nnh has been accurately measured in only one prestellar source, L1544, {where strong  levels of fractionation, with depletion in $^{15}$N,} are found ($^{14}$N/$^{15}$N$\: \approx 1000$). In this paper we extend the sample to three more \textit{bona fide} prestellar cores, in order to understand if the antifractionation in \nnh is a common feature of this kind of sources. }
   {We observed \nnh, \nqn and \qnn in L183, L429 and L694-2 with the IRAM 30m telescope. We modeled the emission with a non-local  radiative transfer code in order to obtain accurate estimates of the molecular column densities, including the one for the optically thick \nnh. {We used the most recent collisional rate coefficients available, and with these we also re-analysed the L1544 spectra previously published.}}
   {The obtained isotopic ratios are in the range $630-770$ and significantly differ with the value, predicted by the most recent chemical models, of  {$\approx 440$}, close to the protosolar value. Our prestellar core sample shows high level of depletion of $^{15}$N in diazenylium, as previously found in L1544. A revision of the N chemical networks is needed in order to explain these results.}
   {}


   \maketitle
%

\section{Introduction}
In the last two decades the chemistry of nitrogen, the fifth most abundant element in the Universe, has raised interest in the context of understanding the formation of interstellar materials and of our own Solar System. In particular, the isotopic ratio \ratio  seems to represent an important diagnostic tool to follow the evolutionary process from the primitive Solar Nebula (where  measurements indicate $^{14}\text{N}/^{15}\text{N}\approx440-450$, \citealt{Marty11, Fouchet04}) up to present. The materials of the Solar System, from meteorites to the Earth's atmosphere, are enriched in $^{15}$N, with the exception of Jupiter atmosphere. Measurements of N$_2$ in the terrestrial atmosphere led to the result of \ratio$\approx 272$ {\citep{Nier50}}, and {carbonaceous} chondrites show isotopic ratios as low as 50 \citep{Bonal10}, suggesting that at the origin of the Solar System multiple nitrogen reservoirs were present \citep{Hily-Blant17}. \par
The nitrogen isotopic ratio has been measured also in different cold environments of the interstellar medium (ISM), and the results show a remarkable spread. \cite{Gerin09} found \ratio$=350-810$ in NH$_3$ in a sample of low mass dense cores and protostars, while using HCN, \cite{Hily-Blant13} found \ratio$=140-360$ in prestellar cores\footnote{The reader must be aware that isotopic ratios measured using HCN (or HNC) depends on the value assumed for the $^{12}$C/$^{13}$C ratio \citep{Roueff15}.}. Using \nnh, \cite{Bizzocchi13} report a value of \ratio$=1000\pm200$ in L1544. In high-mass star-forming regions, the ratio spans a range from 180 up to 1300 in \nnh \citep{Fontani15}, and from 250 to 650 in HCN and HNC \citep {Colzi18a}\footnote{A summary of the measured isotopic ratios can be found in \cite{Wirstrom16}.}. \par
From the theoretical point of view, these results, and especially the very high ratio of L1544, are difficult to explain. The first chemical models addressing the N-fractionation \citep{Terzieva00} suggested that diazenylium (\nnh) should experience a modest enrichment in $^{15}$N thorough the ion-neutral reactions:
\begin{align}
\label{Nreac1}
\text{N}_2\text{H}^+ + {^{15}\text{N} }& \rightleftharpoons \text{N}^{15} \text{NH}^+ + \text{N} \; ,\\
\label{Nreac2}
\text{N}_2\text{H}^+ + {^{15}\text{N}} & \rightleftharpoons {^{15} \text{N}} \text{NH}^+ + \text{N} \; .
\end{align}
A further development of the chemical network made by Charnley and Rodgers led to the so-called superfractionation theory. According to it, extremely high enhancements in $^{15}$N are expected in molecules such as \nnh or NH$_3$, when CO is highly depleted in the gas phase \citep{CharnleyRodgers02,RodgersCharnley08}. Recently, however, based on ab initio calculations, \cite{Roueff15} suggested that the reactions \eqref{Nreac1} and \eqref{Nreac2} do not occur in the cold environments due to the presence of an entrance barrier. As a consequence, no fractionation is expected and the \ratio ratio in diazenylium should be close the protosolar value of $\approx440$, {which is assumed to be valid in the local ISM, according to the most recent results \citep[e.g.][]{Colzi18a,Colzi18b}}. {This value, however, can be considered as an upper limit, since other recent works suggest a lower value for the elemental N-isotopic ratio in the solar neighborhood (e.g. $^{14}\text{N}/^{15}\text{N}\approx 300$, \citealt{Kahane18}, or $^{14}\text{N}/^{15}\text{N}\approx 330$, \citealt{Hily-Blant17}). None of these values are consistent with the} anti-fractionation seen for instance by \cite{Bizzocchi13}. More recently, \cite{Wirstrom17} included the newest rate coefficients from \cite{Roueff15} in a chemical model that takes into account also spin-state reactions, but their predictions fail in reproducing high depletion levels, as well as the high fractionation measured in HCN and HNC.  \par
So far, the observational evidence of anti-fractionation in low-mass star forming regions has been sparse due to the difficulty of such investigations, which require very long integration times ($\gtrsim 8\,$h). Diazenylium presents a further complication. Often, in fact, the \nnh (1-0) emission is optically thick and presents hyperfine excitation anomalies that deviate from the Local Thermodynamic Equilibrium (LTE) conditions \citep{Daniel06, Daniel13}. Thus, a fully non-LTE radiative transfer approach must be adopted, requiring knowledge of the physical structure of the observed source. This method has been up to now applied to only a few sources at early stages, besides L1544. One is the core Barnard 1b, in which isotopic ratios of $\text{N}_2\text{H}^+ / \text{N}^{15} \text{NH}^+ = 400^{+100}_{-65}$ and  $\text{N}_2\text{H}^+ / ^{15}\text{NNH}^+ >600$ were measured by \cite{Daniel13}. A second study was performed in the L16923E core in L1689N, and resulted in $^{14}\text{N}/^{15}\text{N} = 300^{+170}_{-100}$ \citep{Daniel16}.  These two sources, however, are not truly representative of the prestellar phases. Barnard 1b hosts in fact two extremely young sources with bipolar outflows \citep{Gerin15}. 16293E in turn is located very close to the Class 0 protostar IRAS 16293-2422, and it is slightly warmer than typical prestellar cores ($T_{\text{dust}} = 16\,$K, \citealt{Stark04}). We can thus say that the L1544 \ratio ratio in {\nnh} appears peculiarly high, raising the doubt that it could represent an isolated and pathological case. \par
In this paper, we present the analysis of three more objects: L183, L429, and L694-2. These are all \textit{bona fide} prestellar cores  according to \cite{Crapsi05}, due to their centrally peaked column density profiles and high level of deuteration. As in the case of L1544, we modeled their physical conditions and used a non-LTE code for the radiative transfer of \nnh, \nqn, and \qnn emissions. Our results confirm the depletion of $^{15}$N in diazenylium in this kind of sources. 

\section{Observations}
The observations towards the three prestellar cores L183, L429 and L694-2 were carried out with the Institut de Radioastronomie Millim\'etrique (IRAM) 30m telescope, located at Pico Veleta (Spain), during three different sessions. The telescope pointing was checked frequently on planets (Uranus, Mars, Saturn) or a nearby bright source (W3OH), and was found to be accurate within $4''$.  We used the EMIR receiver in the E090 configuration mode. The tuning frequency for the three observed transitions are listed in Table \ref{Lines}. The hyperfine rest frequencies of \qnn and \nqn were taken from \cite{Dore09}. The single pointing observations were performed using the frequency-switching mode. We used the VESPA backend with a spectral resolution of $20\,$kHz, corresponding to $0.06\,$\kms at $90\,$GHz. We observed simultaneously the vertical and horizontal polarizations, and averaged them to obtain the final spectra. \par
\begin{table}[h]
\renewcommand{\arraystretch}{1.4}
\centering
\caption{Rest frequencies of the observed transitions and $1\sigma$ uncertainties. }
\label{Lines}
\begin{tabular}{ccc}
\hline  
Species & Line  & Frequency (MHz) \\
\hline
\nnh     & $J = 1 \rightarrow 0$ & $93173.3991 \pm 0.0004$\tablefootmark{a}   \\
\nqn     & $J = 1 \rightarrow 0$ & $91205.6953 \pm 0.0006$\tablefootmark{b}   \\
\qnn     & $J = 1 \rightarrow 0$ & $90263.8360 \pm 0.0004$\tablefootmark{b}   \\
\hline
\end{tabular}
\tablefoot{
\tablefoottext{a}{From our calculations based on spectroscopic constants of \cite{Cazzoli12} } \\
\tablefoottext{b}{From our calculations based on spectroscopic constants of \cite{Dore09}}
}
\end{table}
L694-2 was observed in good weather condition during July 2011, integrating for $1.15 \,$h for the \nnh (1-0) transition, for $8.9\,$h for \nqn (1-0), and for $9.0\,$h for \qnn (1-0). L183 was observed in July 2012 in good to excellent weather conditions. The total integration times were $11.25\,$min ($\mathrm{N_2H^+}$) and $4.3\,$h ($\mathrm{N^{15}NH^+}$). L429 was observed during two different sessions (July 2012 and July 2017) in average weather conditions. We integrated for a total of $23 \,$min for \nnh (1-0) and $5.7 \,$h on \nqn (1-0). We also observed for $1.2\,$h at the \qnn (1-0) frequency, but we did not detect any signal. For all the sources, we pointed at the millimetre dust peak \citep{Crapsi05}. The core coordinates, together with their distances and locations, are summarised in Table \ref{sources}.  \par
\begin{figure}[h]
\centering
\includegraphics[scale = 0.13]{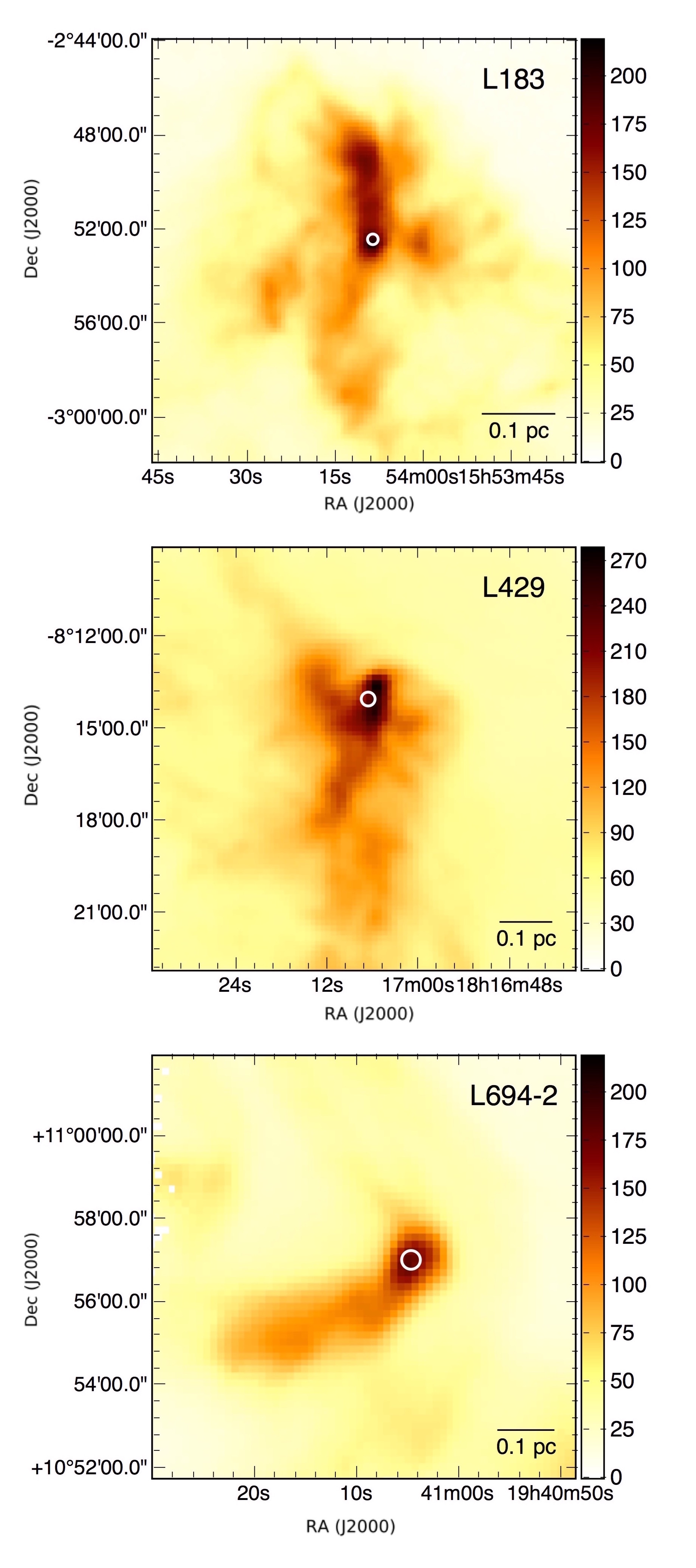}
\caption{The three prestellar cores as seen in dust thermal emission at $350\, \mu$m with Herschel SPIRE camera, in units of MJy$\,$sr$^{-1}$. From top to bottom: L183, L429, L694-2. The scalebar is indicated in each bottom-right corner. The white circles represent the positions of the IRAM pointings and the size of the beam. \label{Cores350}}

\end{figure}

Complementary Herschel SPIRE data, used to obtain the density maps of the sources (see Sec. \ref{PhysMod}), were taken from the \textit{Herschel Science Archive}. The observation ID are: 1342203075 (L183), 1342239787 (L429), and 1342230846 (L694-2). We selected the highest processing level data, already zero-point calibrated and imaged (SPG version: v14.1.0). Figure \ref{Cores350} shows the three cores as seen with the Herschel SPIRE instrument at $350\, \mu$m, as well as the positions of the single-pointing observations performed with IRAM.

\begin{table*}
\renewcommand{\arraystretch}{1.4}
\centering
\caption{Sources' coordinates, distances and locations.}
\label{sources}
\begin{tabular}{cccc}
\hline
Source & Coordinates\tablefootmark{a}                                & Distance (pc)\tablefootmark{b} & Location        \\
\hline
L183   & 15$^h$54$^m$8.32$^s$, -2$^{\degree}$52$'$23.0$''$ & 110           & High lat. cloud    \\
L429   & 18$^h$17$^m$6.40$^s$, -8$^{\degree}$14$'$0.0$''$  & 200           & Aquila Rift     \\
L694-2 & 19$^h$41$^m$4.50$^s$, 10$^{\degree}$57$'$2.0$''$  & 250           & Isolated core \\
\hline
\end{tabular}
\tablefoot{
\tablefoottext{a}{Coordinates are expressed as RA, Dec (J2000)} \\
\tablefoottext{b}{Distances taken from: \cite{Crapsi05} (L429, L694-2), \cite{Pagani04} (L183). }
}
\end{table*}
\section{Results}
\begin{figure*}[h]
\centering
\includegraphics[scale = 0.1]{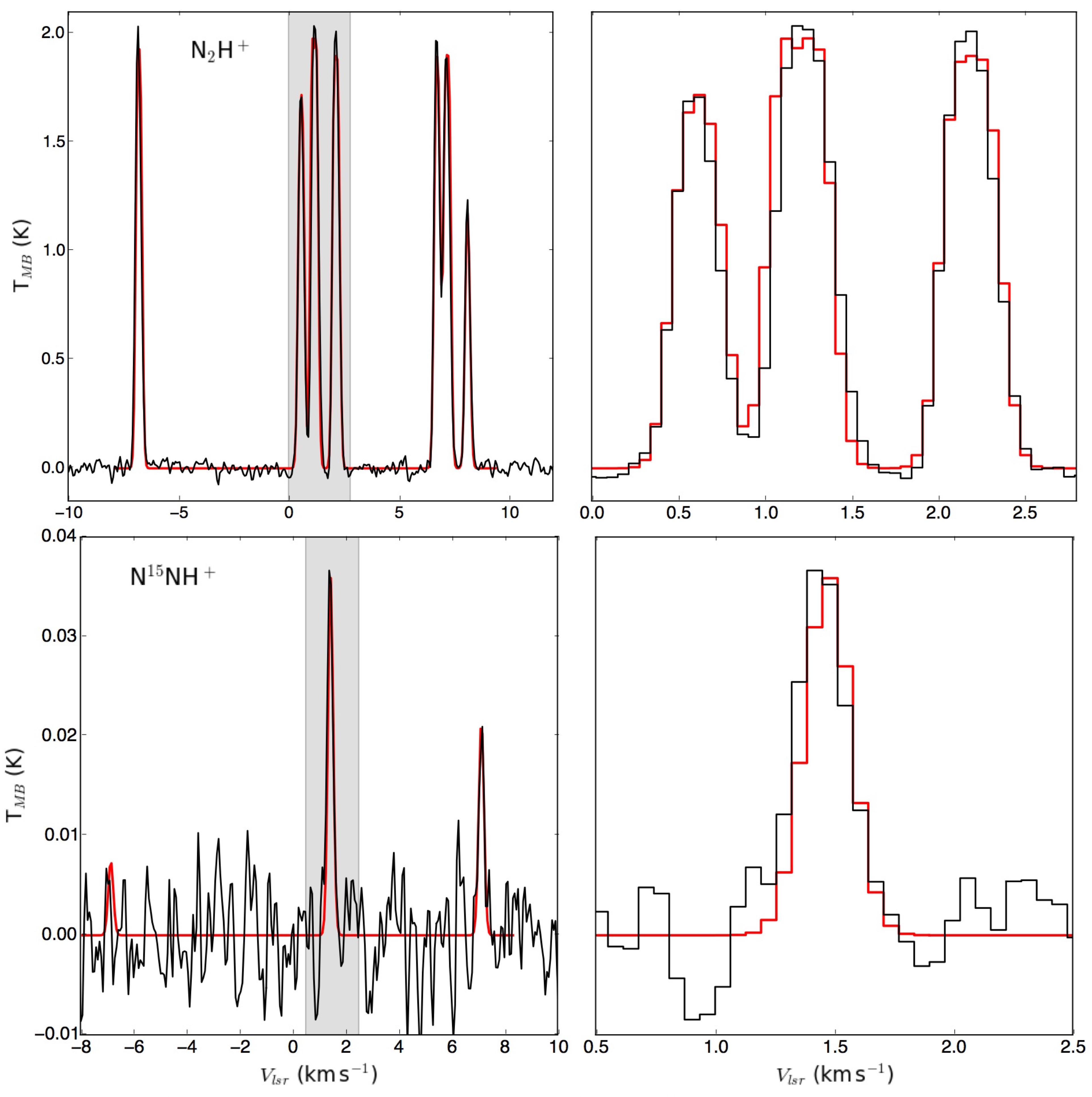}
\caption{Observed spectra (black) and modeled ones (red) in L183, for \nnh (top) and \nqn (bottom). The modeling was performed with MOLLIE as described in Sec. \ref{Analysis}. The left panels show the entire acquired spectra, while the right ones are zoom-in of the grey shaded velocity range. \label{L183}}
\end{figure*}

\begin{figure*}[h]
\centering
\includegraphics[scale = 0.18]{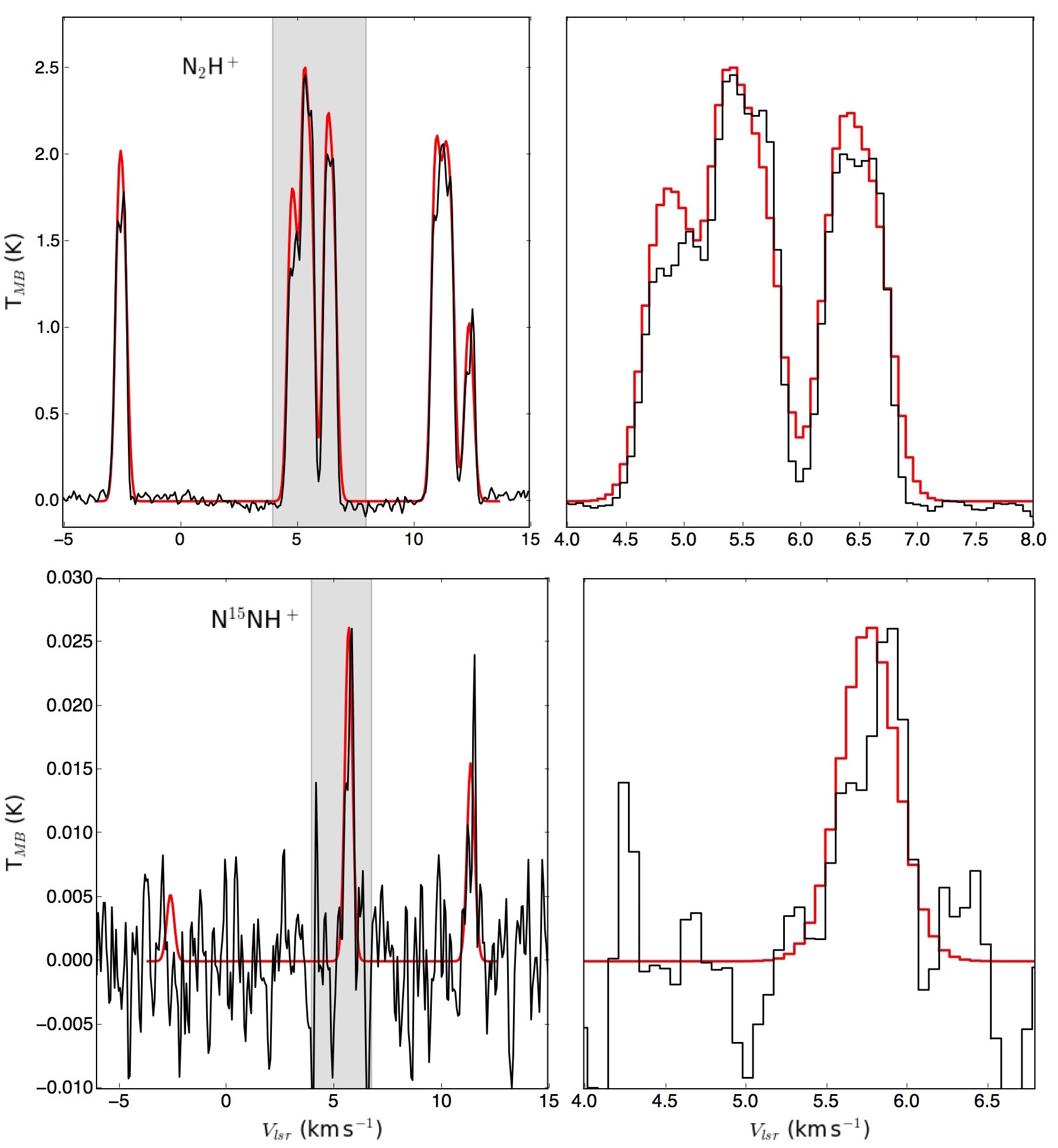}
\caption{Observed spectra (black) and modeled ones (red) in L429, for \nnh (top) and \nqn (bottom). The modeling was performed with MOLLIE as described in Sec. \ref{Analysis}{, and includes the infall velocity profile}. The left panels show the entire acquired spectra, while the right ones are zoom-in of the grey shaded velocity range. \label{L429}}
\end{figure*}

\begin{figure*}[h]
\centering
\includegraphics[scale = 0.17]{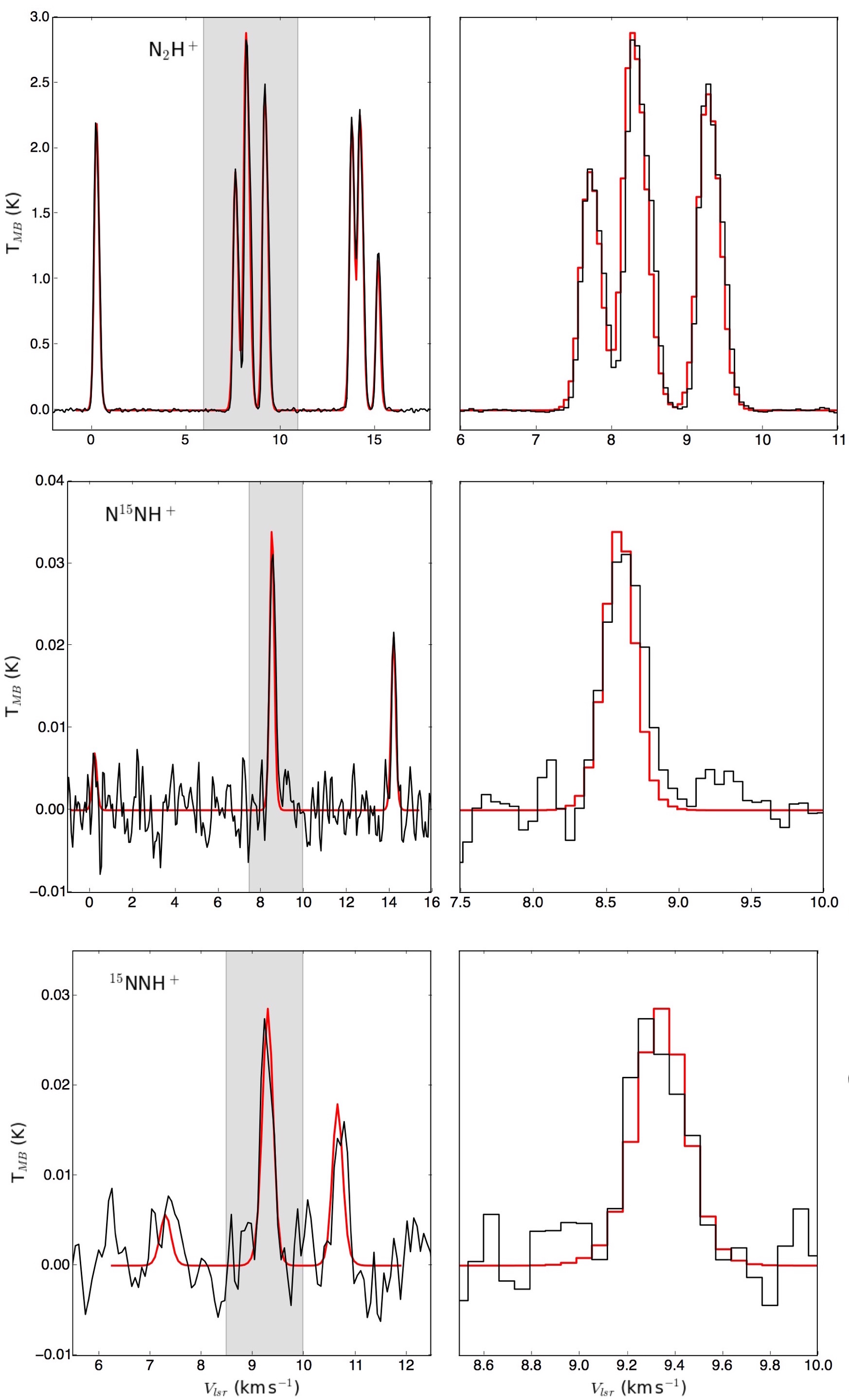}
\caption{Observed spectra (black) and modeled ones (red) in L694-2, for \nnh (top panel), \nqn (middle panel), and \qnn (bottom panel). The modeling was performed with MOLLIE as described in Sec. \ref{Analysis}. The left panels show the entire acquired spectra, while the right ones are zoom-in of the grey shaded velocity range. \label{L694-2}}
\end{figure*}

The obtained spectra are shown in the left panels of Figure \ref{L183} (L183), Figure \ref{L429} (L429), and Figure \ref{L694-2} (L694-2). The data were processed using the GILDAS\footnote{Available at \url{http://www.iram.fr/IRAMFR/GILDAS/}.} software, and calibrated in main beam temperature $T_{\text{MB}}$ using the telescope efficiencies ($F_{\text{eff}}=0.95$ and $\eta_{\text{MB}}=0.80$, respectively) at the observed frequencies. The typical rms is $10-20\,$mK for \nnh(1-0), and $3-4 \,$mK for the spectra of the rarer isotopologues, resulting in good to high-quality detections. The minimum signal-to-noise ratio (SNR) is $\approx 8$, whilst the maximum is $\approx150$. \par
The CLASS package of GILDAS was first used to spectrally fit the data. We used the HFS fitting routine, which models the hyperfine structure of the analyzed transition assuming local thermodynamic equilibrium (LTE).  Especially in the case of the \nnh(1-0) line, this routine is not able to reproduce the observed data, due to the fact that the LTE conditions are not fulfilled. This is not due only to optical depth effects, which are taken into account in CLASS routines, but also to the fact that the excitation temperature is not the same for all the hyperfine transitions. A more refined approach that uses non-LTE analysis is therefore needed to compute reliable column densities, and will be discussed later (see Sec. \ref{Analysis}). Nevertheless, the CLASS analysis provides reliable results for the local standard of rest velocity ($V_{\text{LSR}}$), whose values are summarized in Table \ref{Vlsr}. Different isotopologues give generally consistent results, within $3\sigma$, for each source. The values derived from \nnh (1-0) are also in agreement with the literature ones \citep{Crapsi05}. Table \ref{Vlsr} summarizes also the total linewidth (FWHM) obtained with the CLASS fitting routine. 

\begin{table}[!h]
\renewcommand{\arraystretch}{1.4}
\centering
\footnotesize
\caption{$V_{\text{LSR}}$ estimated from the CLASS HFS fitting routine.}
\label{Vlsr}
\begin{tabular}{cccc}
\hline
Source                  & Line       & $V_{\text{LSR}}$ (\kms) & FWHM (\kms) \\
\hline
                        
\multirow{2}{*}{L183}   & \nnh (1-0) & $2.4145\pm0.0004$ &   $0.222\pm 0.001$       \\
                        & \nqn (1-0) & $2.390\pm0.009$  &   $0.28 \pm 0.03$     \\
                        \hline
 \multirow{2}{*}{L429}   & \nnh (1-0) & $6.7141\pm 0.0006$  & $0.401 \pm 0.001$       \\
               & \nqn (1-0) & $6.77\pm0.02$ &    $ 0.41\pm0.06$      \\
               \hline 
\multirow{3}{*}{L694-2} & \nnh (1-0) & $9.5577\pm 0.00014$  &     $0.2635 \pm 0.0004  $ \\
                        & \nqn (1-0) & $9.562\pm0.007$  &  $0.32 \pm 0.03$   \\
                        & \qnn (1-0) & $9.563\pm0.011$  &   $0.30 \pm 0.02$     \\
                        \hline                   
\end{tabular}
\end{table}

\section{Analysis \label{Analysis}}
Our aim is to derive the column density of the different isotopologues, and to compute from their ratios the values of the corresponding $\mathrm{^{14}N/^{15}N}$, assuming that they are tracing the same regions. For the previously mentioned reasons, this is not possible using a standard LTE analysis (such as the one presented in the Appendix A of \citealt{Caselli02b}), as already shown for instance in the analysis of L1544 by \cite{Bizzocchi13}. Therefore, we used a non-LTE method, based on the radiative transfer code MOLLIE \citep{Keto90, Keto04}. MOLLIE can produce synthetic spectra of a molecule arising from a source with a given physical model. In particular, it is able to treat the case of overlapping transitions, and thus can properly model the crowded \nnh (1-0) pattern. In what follows, we first describe the construction of the core models, and then present the analysis of the observed spectra using MOLLIE. {The case of L429, which presents peculiar issues, is treated separately.}
\subsection{Source physical models \label{PhysMod}}
MOLLIE is able to treat genuine 3D source models. Nevertheless, for the sake of simplicity, we chose to model the cores in our sample as spherically symmetric (1D). As one can see in Figure \ref{Cores350}, this assumption holds reasonably well for the densest parts {of all cores}\footnote{L694-2 was modeled as {an} elongated cylinder with the axis almost along the line of sight by \cite{Harvey03b,Harvey03a}, but for the sake of simplicity, and given the { relative roundness of the} source at high density (shown in Figure \ref{Cores350}), we adopted a 1D model.}.  For L183, a more sophisticated{, 2D model,} have already been developed in our team (Lattanzi et al., in prep.). {This consists of a cylinder, with the axis lying on the plane of sky.} In order to be consistent with the analysis of the other two cores, we decided to average this model in concentric annuli {on the plane of sky} to obtain a one-dimensional profile.\par
The simplest 1D model consists of a volume density radial profile and a temperature {radial} profile. We thus assume that the gas kinetic temperature and the dust temperature are equal. This is strictly true only when gas and dust are coupled ($n>10^{4-5} \,$cm$^{-3}$, \citealt{Goldsmith01}), but we do not have for all the sources enough information on the spatial distribution of the gas temperature, which would require maps of NH$_3$ (1,1) and (2,2) with JVLA (see \citealt{Crapsi07}). On the other hand, the available continuum data allow us to determine reliable values for the dust temperature with a resolution of $\approx 40''$. \par
The volume density profile is derived from the analysis of the Herschel SPIRE maps at $250\, \mu$m, $350\, \mu$m, and $500\, \mu$m, as follows. Since we are interested in the core properties, we filtered out the contribution of the diffuse, surrounding material with a background subtraction. We computed the average flux of each map in the surrounding of the cores, at a distance of $\approx500-800''$. This was assumed to be the background contribution, and was subtracted from the SPIRE images pixel by pixel. Then, the background-subtracted SPIRE maps were fitted {simultaneously} using a modified black body emission, in order to obtain the dust column density map of the source {(for a complete description of the procedure, see for instance Appendix B of \citealt{Redaelli17})}. We adopted the optically thin approximation, and a gas-to-dust ratio of 100 \citep{Hildebrand83} to derive the H$_2$ column density. The dust opacity is assumed to scale with the frequency as
\begin{equation}
\kappa_{\nu} = \kappa_{250\mu\text{m}} \left( \frac{\nu}{\nu_{250\mu\text{m}}}\right )^{\beta} \: , 
\end{equation}
where {$\kappa_{250\mu\text{m}} = 0.1 \,$cm$^{2}\,$g$^{-1}$} is the reference value at  $250\, \mu$m \citep{Hildebrand83}, and $\beta = 2.0$, a suitable value for low-mass star-forming regions \citep{Chen16, Chacon-Tanarro17, Bracco17}. {From this procedure one also gets the line-of-sight averaged dust temperature map of each source. These data, however, were not used in the following analysis.} \par
{The obtained column density map was averaged in concentric annuli starting from the densest pixel, and then a Plummer profile was fitted to the obtained points according to:}
\begin{equation}
N(r) = \frac{N(\text{H}_2)_{\text{peak}}}{\left[ 1+ \left( \frac{r}{r_0}\right)^2\right]^{\frac{p-1}{2}}} \; .
\end{equation}
{The obtained best-fit values of the free parameters (the characteristic radius $r_0$, the power-law parameter $p$ and the central column density $N(\text{H}_2)_{\text{peak}}$) can be used to derive the volume density profile $n(r)$, according to \cite{Arzoumanian11}, following:}
\begin{equation}
n(r) = \frac{n_0}{\left[ 1+ \left( \frac{r}{r_0}\right)^2\right]^{\frac{p}{2}}} \; .
\end{equation}
Table \ref{PlummerPara} summarizes the best fit values of the Plummer-profile fitting of each source. The values obtained for the $p$ exponent are in the range $\approx 2-3.5$, quite consistent with those found for other cores using similar power-law profile shapes \citep[e.g. in][]{Tafalla04, Pagani07}. The profiles obtained with this method typically show $n_0 \lesssim \, \text{a few} \, 10^5\,$cm$^{-3}$, and fall below $10^5\,$cm$^{-3}$ within the central $\approx 3500\,$AU. They thus fail to reach the high volume densities typical of prestellar cores centres. {In fact, the integrated $n(r)$ profiles along the line of sight result in column density values lower by a factor of 2-4 compared to the results of \cite{Crapsi05}, although the dust opacity value used in that work is consistent within 15\% with ours. This is} due to the poor angular resolution of the SPIRE maps, which were all convolved to the beam size of the $500\, \mu$m map ($\approx 38''$). The central regions of dense cores are in fact better traced with millimetre dust emission observations performed with large telescopes, which allow to better see their cold and concentrated structure. In order to correct for this, we artificially increased the density in the central part ($5-10\%$ of the total core radius $r_c$), until the column density derived from this profile is consistent with the value obtained from $1.2\,$mm observations. The inserted density profile was taken from the central part of the profile developed through hydrodynamical simulations for L1544 in \cite{Keto15}, a model known to work well to reproduce the prestellar core properties. \par
\begin{figure}[h]
\centering
\includegraphics[scale = 0.1]{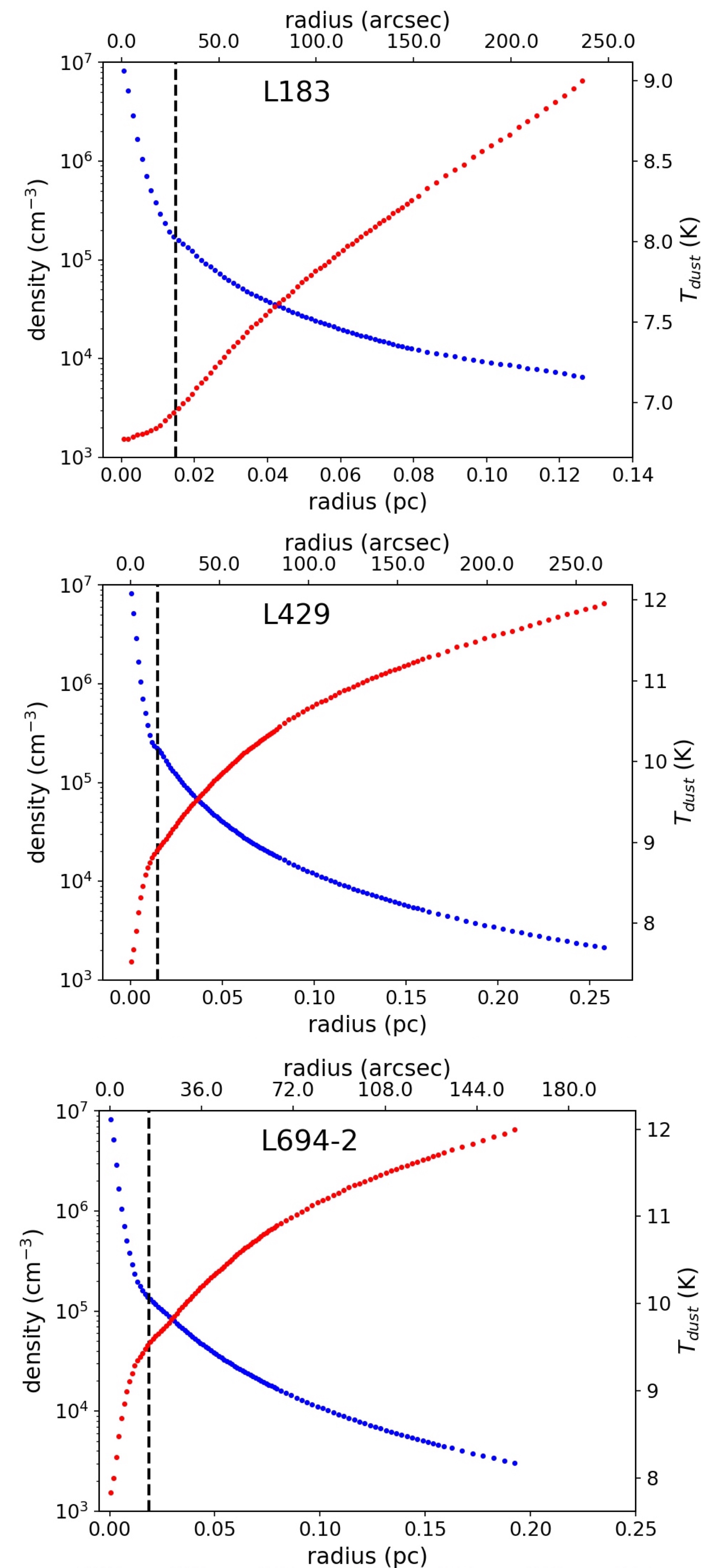}
\caption{The volume density profile (blue dots) and the dust temperature profile (red dots) for the three cores (from top to bottom: L183, L429, L694-2), {as a function radius in both pc and arcsec}. The vertical, dashed lines represent the radius within which the density was artificially increased (see text for details). \label{CoresModels}}
\end{figure}
The volume density profile derived in the previous paragraph is used as an input for the Continuum Radiative Transfer code CRT \citep{Juvela01, Juvela05} to derive the dust temperature ($T_{\text{dust}}$) profile. The CRT is a Monte Carlo code that computes the emerging emission and the dust temperature, given a background radiation field. For the latter, we used the standard interstellar radiation field of \cite{Black94}. {Since we want to model cores embedded in a parental cloud, the background radiation field has to be attenuated. Our team has the tabulated values for the \cite{Black94} model with an attenuation corresponding to a visual extinction of $A_V=1, \; 2, \; \text{and} \;10\,$mag. We tested all the three options, and found that generally the first two provide too warm temperatures. We thus decided to assume that radiation impinging on the cores is attenuated by an ambient cloud, whose thickness corresponds to a visual extinction of $A_V=10\,$mag.  The external radiation field can still be multiplied by a factor $k$, in order to correctly reproduce the emitted surface brightness. To determine this parameter, we tested a number of values in the range $k=0.5- 5.0$, and for each one we computed the synthetic flux emitted at the SPIRE wavelengths at the cores' centres. We adopted the model that provides the best agreement with the observations\footnote{When the observations were available, we also simulated the flux at mm wavelengths.}. Typically, we needed to increase the external radiation field by a factor of 2-3, suggesting that the assumed thickness of the ambient cloud was too large, and that the real, correct attenuation is somewhat in between $2$ and $10\,$mag. This assumption is reasonable, as the H$_2$ column density derived from the SPIRE maps around the cores is usually $\mathrm{\approx 5 \,10^{21}}\,$cm$^{-2}$, albeit we do not know the three-dimensional structure of the cloud.} The dust opacities are taken from \cite{OssenkopfHenning94} for unprocessed dust grains covered by thin icy mantles. This choice is made so that the dust opacity values used in this part are consistent with the one from \cite{Hildebrand83}, used for fitting the SPIRE maps. \par
The models built so far are static, i.e. the velocity field is zero everywhere. However, we know that many cores show hints of infall or expansion motions \citep{Lee01}. The velocity field can heavily impact the spectral features, and, when possible, it must be taken into account. For L183, in Sec. \ref{MOLLIE} we will show that the static model is good enough to properly reproduce the observations. For L694-2, we used the infall profile derived in \cite{Lee07} using high spatial resolution HCN data. L429 represents a more difficult case, and will be accurately treated in the next subsections.
\begin{table}[h]
\renewcommand{\arraystretch}{1.4}
\centering
\caption{Summary of the best fit values for the parameters of the Plummer profiles, for each source.}
\label{PlummerPara}
\begin{tabular}{ccc}
\hline
$n_0\, (\text{cm}^{-3})$ & $p$         & $r_0 ('')$ \\
\hline
\multicolumn{3}{c}{L183}                              \\
$(4.1 \pm0.7)10^5$       & $1.95\pm0.10$ & $29\pm 2$  \\
\hline
\multicolumn{3}{c}{L429}                              \\
$(4.1 \pm0.7)10^5$       & $1.86\pm0.08$ & $16 \pm1$  \\
\hline
\multicolumn{3}{c}{L694-2}                            \\
$(1.8 \pm0.5)10^5$       & $3.32\pm0.56$ & $41\pm6$ \\
\hline 
\end{tabular}
\end{table}

\subsection{Spectral modeling with MOLLIE \label{MOLLIE}}
The physical models developed in Sec. \ref{PhysMod} are used as inputs for MOLLIE. The structure of each core is modeled with three nested grids of increasing resolutions towards the centre, each composed by 48 cells, onto which the physical quantities profiles are interpolated. The collisional coefficients used are from \cite{Lique15}, who computed them for the main isotopologue and the most abundant collisional partner, $p$-H$_2$.  The $o$-H$_2$ is thus neglected, which is a reasonable assumption given that the orto-to-para ratio (OPR) in dense cores is expected to be very low (in L1544, $\text{OPR}\approx 10^{-3}$, \citealt{Kong15}). The collisional coefficient for \nqn and \qnn have been derived from those of \nnh following the method described in Appendix \ref{HypRates}. \par
Our fit procedure has two free parameters, the turbulent velocity dispersion, $\sigma_{\text{turb}}$, and the molecular abundance $X_{\text{st}}$ with respect to H$_2$, assumed to be radially constant. Since MOLLIE requires very long computational times to fully sample the parameters space and find the best-fit values, we proceeded with a limited parameter space sampling. We first set the $\sigma_{turb}$ value, testing $\approx 5$ values on the \nnh (1-0) spectra. This value is kept fixed also for the \nqn and \qnn (1-0) lines. Then we produced $8$ synthetic spectra for each transition, varying each time the initial abundance, and convolving them to the $27''$ IRAM beam. The results were compared to the observations using a simple $\chi^2$ analysis, i.e. computing:
\begin{equation}
\chi^2 = \sum_i \left\{ \frac{\left( T_{\text{MB,obs}}^i - T_{\text{MB,mod}}^i \right)^2}{\sigma_{\text{obs}}^2} \right\} \: ,
\label{chi2}
\end{equation}
where $T_{\text{MB,obs}}^i$ and $T_{\text{MB,mod}}^i$ are the main beam temperature in the $i$-th velocity channel for the observed spectrum and the modeled one, respectively, and $\sigma_{\text{obs}}$ is the rms of the observations. The sum is computed excluding signal-free channels. In order to evaluate the uncertainties, we fitted a polynomial function to the $\chi^2$ distribution and set the lower/upper limits on $X_{st}$ according to $\chi^2$ variations of $\approx 20 \%$ for the \nnh (1-0) spectra and $\approx 15 \%$ for the other isotopologues. We chose these two different limits due to different opacity effects. In fact, the \nnh (1-0) lines are optically thick, and thus changes in the molecular abundance lead to smaller changes in the resulting spectra compared to the optically thin \qnn and \nqn lines. Since the $\chi^2$ distribution is usually asymmetric, so are the error bars. In order to evaluate the column densities, we integrated the product $n(\text{H}_2) \cdot X_{st}$ (convolved to the IRAM beam) along the line of sight crossing the centre of the model sphere. {In Appendix \ref{chi2App}, we report the curves for the $\chi^2$ in the analysed sources.} \par
Figure \ref{L183} and \ref{L694-2} show the best fit spectra (in red), obtained as just described, in comparison with the observed ones (black curve) for L183 and L694-2 , respectively. The overall agreement is good, and most of the spectral features are well reproduced, as seen in the right panels, which show a zoom-in of the main component. 
\subsection{The analysis of L429}
L429 represents a more difficult  case to model. As one can see in Figure \ref{L429} and from the last column of Table \ref{Vlsr}, the \nnh (1-0) line is almost a factor of two broader than in the other two sources. This may be due to the fact that this core is located in a more active environment, the Aquila Rift, but can also be a hint of multiple components along the line of sight. Moreover, concerning its velocity field, \cite{Lee01} listed L429 among the ``strong infall candidates'' while in \cite{Sohn07} the analyzed HCN spectra shows both infall and expansion features. A full charachterization of the dynamical state of the source and its velocity profile would require high quality, high spatial resolution maps of molecular emission, which is beyond the scope of this paper. At the first stage we tried to fit the observed spectra first increasing $\sigma_{\text{turb}}$. The static model is however unable to reproduce the hyperfine intensity ratios, and thus we adopted the infall profile of L694-2. The agreement with the observations increased significantly, meaning that a velocity field is indeed required to model the spectra. Due to the difficulties in analyzing this source, the $\chi^2$ analysis previously described is not suitable, {because it presents an irregular shape and its minimum corresponds to a clearly wrong solution, due to the fact that it is not possible to simultaneously reproduce the intensity of all the hyperfine components}. We therefore determined $X_{\text{st}}$ in the same way as for $\sigma_{\text{turb}}$, testing multiple values. We then associated the uncertainty to this   value using the largest relative uncertainties found in the other two sources ($24\%$ for \nnh (1-0) and $25\%$  for \nqn (1-0)). 
\subsection{Obtained results}

Table \ref{Results} summarizes the values of $\sigma_{\text{turb}}$, $X_{\text{st}}$, and column density $N_{\text{mol}}$ for each line in the observed sample. For a sanity check, since the rare isotopologues transitions are optically thin and do not present intensity anomalies, we derived their molecular densities using the LTE approach of \cite{Caselli02b}, focusing on the main component only.  The results of this analysis are shown in the 6th column of Table \ref{Results} ($N_{\text{mol}}^{\text{LTE}}$). One can note that these values are consistent with the ones derived through the non-LTE method. {The L183 physical structure and \nnh emission have been previously modelled by \cite{Pagani07}. It is interesting to notice that their best fit profiles for both density and temperature are close to ours, even though their model is warmer in the outskirts of the source. Furthermore, despite a different abundance profile, their derived \nnh column density is consistent with our value.} \par

\begin{table*}
\renewcommand{\arraystretch}{1.6}
\centering
\caption{Parameters and results of the modelling with MOLLIE.}
\label{Results}
\begin{tabular}{c|cccccc}
\hline
Source & Line & $\sigma_{\text{turb}}$/\kms & $X_{\text{st}}/10^{-13}$ & $N_{\text{mol}}/10^{10}$cm$^{-2}$  & $N_{\text{mol}}^{\text{LTE}}/10^{10}$cm$^{-2}$  & \ratio \\
\hline
L183 &\nnh     & 0.12                   &   $ 2.50^{+0.25}_{-0.60} \, 10^{3}$       &        $1.29^{+0.13}_{-0.31} \, 10^{3}$  & -&        \\
         &\nqn     & 0.12                   &    $ 3.75^{+0.95}_{-0.75} $      &          $ 1.93^{+0.49}_{-0.38} $  &    $ 2.24^{+0.54}_{-0.54} $ & $ 670^{+150}_{-230}$    \\
\hline
L429  & \nnh     & 0.23                   &   $4.5 \,10^3$     &   $1.82^{+0.44}_{-0.44}\,10^{3}$   & - & \\
	&\nqn     & 0.23                   &   $5.5$       &        $2.5^{+0.63}_{-0.63}$ &  $ 2.46^{+0.44}_{-0.44} $  &      $730^{+250}_{-250}$      \\
	 \hline
L694-2 & \nnh     & 0.12                   &    $ 3.50^{+0.60}_{-0.35} \, 10^{3}$       &          $ 1.26^{+0.22}_{-0.12} \, 10^{3}$   & - &          \\
	& \nqn     & 0.12                   &    $ 6.00^{+1.00}_{-1.00} $     &   $ 2.17^{+0.36}_{-0.36} $   &    $ 2.74^{+0.43}_{-0.43} $ &  $ 580^{+140}_{-110}$    \\
	& \qnn     & 0.12                   &      $ 5.00^{+0.90}_{-1.20} $      &        $ 1.81^{+0.32}_{-0.44} $   &  $ 2.13^{+0.37}_{-0.37} $ & $   700^{+210}_{-140}$   \\
\hline
L1544\tablefootmark{a} & \nnh     &0.075               &    $ 5.50^{+1.25}_{-0.75} \, 10^{3}$       &          $ 1.73^{+0.39}_{-0.24} \, 10^{3}$   &- &            \\
	& \nqn     & 0.075                  &    $6.00^{+1.00}_{-1.40} $     &          $ 1.89^{+0.31}_{-0.44} $ & $ 2.45^{+0.57}_{-0.57} $  & $ 920^{+300}_{-200}$    \\
	& \qnn     & 0.075               &     $ 5.50^{+0.95}_{-0.70} $      &       $ 1.73^{+0.30}_{-0.22} $  &  $ 2.02^{+0.28}_{-0.28} $& $ 1000^{+260}_{-220}$   \\
\hline
\end{tabular}
\tablefoot{\tablefoottext{a}{The values for L1544 are based on the data shown in \cite{Bizzocchi13}. The non-LTE modeling uses the updated collisional rates, while the LTE results were derived adopting revised excitation temperature values.}}
\end{table*}
With the values for the molecular column densities found with the fully non-LTE analysis, we can infer the isotopic ratio dividing the main isotolopogue column densities for the corresponding rare isotopologues ones. Uncertainties are propagated using standard error calculation. The results are summarized in the last column of Table \ref{Results}.

\section{Discussion}
\begin{figure}[h]
\centering
\includegraphics[scale = 0.5]{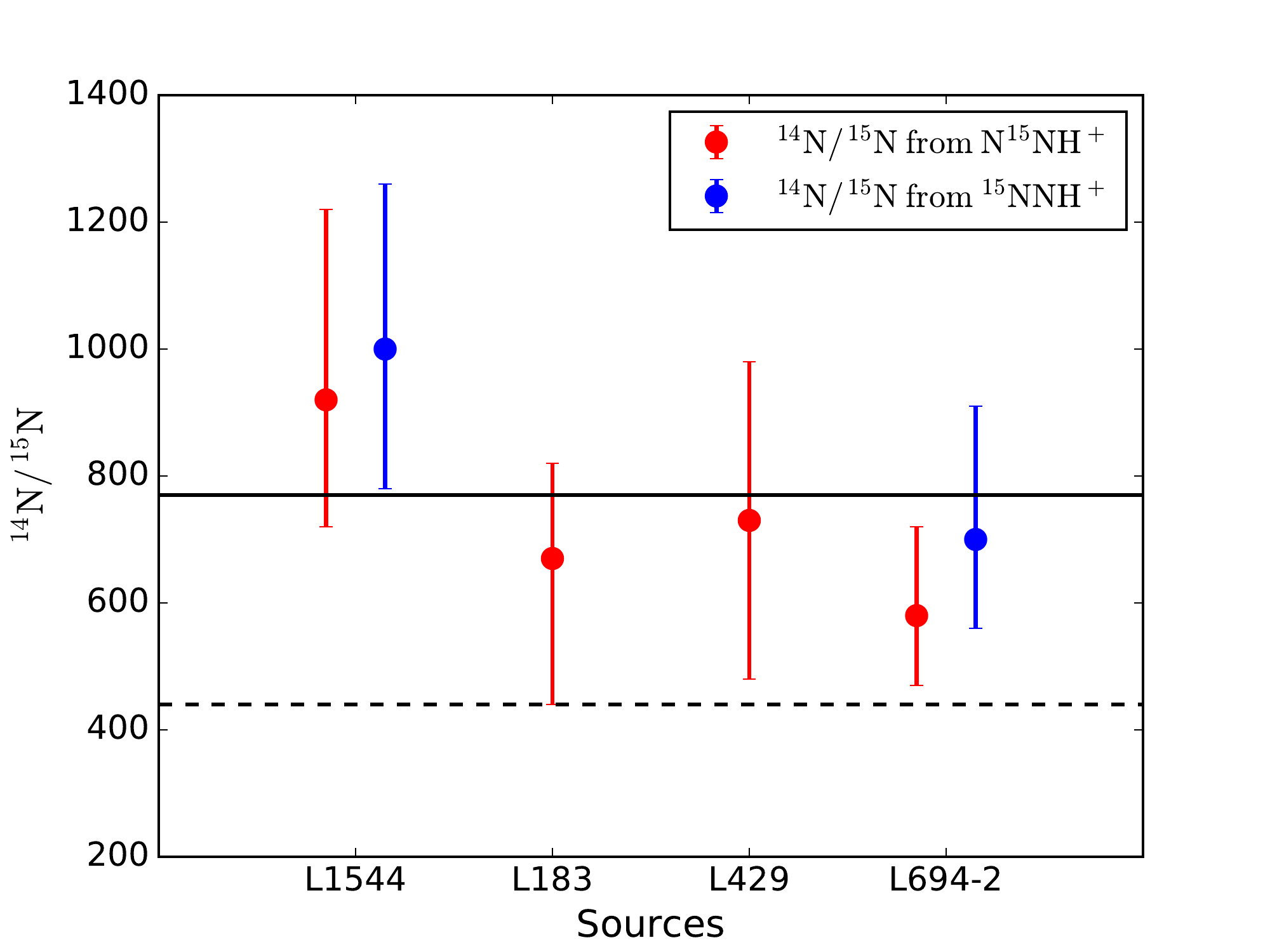}
\caption{The \ratio values obtained in the sample presented in this paper and re-computed for L1544 with errorbars, determined with the method described in the main text.. Red points refer to measurements of \nqn, while blue ones of \qnn. The solid line represents the average value found in the whole sample ($= 770$), while the dashed curve is the protosolar nebula value (440).  \label{Ratios}}
\end{figure}

Figure \ref{Ratios} shows a summary of the obtained isotopic ratios. Since from the analysis of \cite{Bizzocchi13} the collisional rates for the \nnh system have changed, we re-modeled the literature data for this source. The new results are:  $^{14}\text{N}/^{15}\text{N} = 920^{+300}_{-200}$ (using $\mathrm{N^{15}NH^+}$) and $^{14}\text{N}/^{15}\text{N} = 1000^{+260}_{-220}$ (using $\mathrm{^{15}NNH^+}$). They are also shown in Figure \ref{Ratios}. 
These values are perfectly consistent with the already mentioned literature value of $^{14}\text{N}/^{15}\text{N} = 1000 \pm 200$ of \cite{Bizzocchi13}. More recently, \cite{DeSimone18} computed again the nitrogen isotopic ratio from diazenylium in L1544, and found  $^{14}\text{N}/^{15}\text{N} = 228-408$, a result inconsistent with ours. The IRAM data analysed by those authors, though, have a spectral resolution of $50\,$KHz (more than twice coarse than ours). Furthermore, the authors used a standard LTE analysis, which is not suitable for this case, as already mentioned. This point has been studied in detail in \cite{Daniel06, Daniel13}, where the authors showed that in typical core conditions ($T \approx 10\,$K, $n \approx 10^{4-6}\,$cm$^{-3}$) the hypothesis of identical excitation temperature for all hyperfine components in the \nnh (1-0) transition is not valid. Due to radiative trapping effects, in fact, the hyperfines intensities ratios deviates from the ones predicted with LTE calculations.  \par
Figure \ref{Ratios} shows that the computed values are consistent within the source sample. {Due to the large uncertainties, we can conclude that the isotopic ratios are only marginally inconsistent with the value of 440, representative of the protosolar nebula (black, dashed curve). Nevertheless, the trend is clear. Despite the fact that L1544 still present the highest values in the sample, its case is now clearly not an isolated and pathological one. This larger statistics thus supports the hypothesis that diazenylium is $^{15}$N-depleted in cold prestellar cores.} Instead of "super-fractionation" predicted by some chemistry models, \nnh seems to experience "anti-fractionation" in these objects. As already stressed out, this trend cannot be understood within the frame of current chemical models. \cite{Roueff15} predict that the \ratio should be close to the protosolar value  ($\approx 400$). \cite{Wirstrom17} came to very similar conclusions. In both chemical networks, the physical model assumes for the gas a temperature of $10\,$K, which can be up to 40\% higher than the values found for the central parts of the cores ($6-7$\,K). However, further calculations have shown that lowering the temperature by $2-3\,$K does not produce significant differences in the results (Roueff, private communications). {\cite{Visser18} highlighted how isotope-selective photodissociation is a key mechanism to understand the nitrogen isotopic chemistry in protoplanetary disks. We can speculate that different levels of selective photodissociation in different environments could reproduce the variety of N-isotopic ratios that are observed. It will be indeed worthy to further investigate this point by both observational and theoretical point of view.} \par
From the L1544 and L694-2 results, there seems to be a tentative evidence that \nqn is more abundant than \qnn. This can be explained by the theory, according to which the proton transfer reaction 
\begin{equation}
^{15}\text{NN}\text{H}^+ + {\text{N}^{15}\text{N} } \rightleftharpoons \text{N}^{15} \text{NH}^+ +{\text{N}^{15}\text{N} } \; , \\
\end{equation}
tends to shift the relative abundance of the two isotopologues, slightly favouring \nqn due to the fact that \nqn  zero-point energy is lower than the one of \qnn by $8.1\,$K \citep[see reaction RF2 in][]{Wirstrom17}. {It is interesting that the same trend is found also in a very different environment such as OMC-2 FIR4, a young protocluster hosting several protostellar object. In this source, \cite{Kahane18} measured lower values for the isotopic ratio, but in agreement with our result found that \qnn is less abundant than $\mathrm{N^{15}NH^+}$.} However, we emphasise that higher quality and higher statistics observations are needed to confirm this point.

\section{Conclusions}
We have analyzed the diazenylium isotopologues spectra in three prestellar cores, L183, L429 and L694-2 in order to derive nitrogen isotopic ratios. Since LTE conditions are not fulfilled, especially for the \nnh (1-0) transition, we have used a fully non-LTE radiative transfer approach, implemented in the numerical code MOLLIE. We have carefully derived the physical models of the sources, computing their volume density and dust temperature profiles. With these, we were able to produce synthetic spectra to be compared with our observations, in order to derive the best-fit values for the molecular abundances and column densities. Using the same method, we have also re-computed the isotopic ratio of L1544. The difference with the literature value of \cite{Bizzocchi13}, due to changes in the molecular collisional rates, is well within the uncertainties. \par
In our sample of 4 cores, we derived \ratio values in the range $630-1000$. Within the confidence range given by our uncertainties estimation, all our results are inconsistent with the value $\approx 400$, predicted by the current theoretical models. L1544 still presents higher depletion levels than the other sources, but in general all the cores are anti-fractionated. The theoretical bases of such a trend are at the moment not understood. A deep revision of our knowledge of the nitrogen chemistry is required in order to understand the chemical pathways that lead to so low abundances of \nqn and \qnn compared to the main isotopologue. 

\begin{acknowledgements}
We thank the anonymous referee, whose comments helped improving the quality of the manuscript.
\end{acknowledgements}


\appendix
\onecolumn
\section{New hyperfine rate coefficients for the N$_2$H$^+$/$p$-H$_2$ collisional system}
\label{HypRates}
Recently, \citet{Lique15} published a new set of theoretically computed
hyperfine rate coefficients of N$_2$H$^+$ ($X^1\Sigma^+$) excited by
$p$-H$_2$ ($j = 0$).
The scattering calculation is based on a high-level ab initio potential energy surface
(PES), from which state-to-state rate coefficients between the low-lying hyperfine levels
were derived for temperatures ranging from 5 to~70\,K\@.
These new results provided the first genuine description of the N$_2$H$^+$/$p$-H$_2$
collisional system, much improving the one based on previously published studies
\citep{Daniel04,Daniel05} which used the He atom as collision partner.
Indeed, the dissimilar polarizability of H$_2$ and He has a sizeable effect on the
long-range electrostatic interaction and produces A marked difference in the
corresponding collision cross sections.
\citet{Lique15} found discrepancies up to a factor of $\sim 3$
(N$_2$H$^+$/$p$-H$_2$ being the larger), thus indicating that the commonly used scaling
factor of 1.37 (based on reduced masses) is not appropriate for these systems. \par
In a previous non-LTE analysis of N$_2$H$^+$ emission in L1544 \citep{Bizzocchi13}, a
$J$-dependent scaling relation based on HCO$^+$--H$_2$ and HCO$^+$--He rate coefficients
was adopted.
This scheme produced factor ranging into 1.4--3.2 interval with an average ratio of
$\sim 2.3$ for $1\leq j \leq 4$,
and thus allowed for a more reliable modelling of the N$_2$H$^+$ collisional excitation
in the ISM\footnote{%
  In order to be consistent with the formalism employed in collision studies
  \citep[e.g.,][]{Daniel05,Lique15}, in this appendix the
  lower-case symbol $j$ is used for the quantum number of the molecule end-over-end
  rotation.}.
However, the newly computed set of genuine N$_2$H$^+$/$p$-H$_2$ collision data clearly
supersedes the one derived through this procedure.
The RT modelling presented in this paper, were thus performed inserting in MOLLIE the
\nnh/$p$-H$_2$, \nqn/$p$-H$_2$, and \qnn/$p$-H$_2$ rate coefficients derived from
\citet{Lique15}. \par
Hyperfine de-excitation rate coefficients for the main isotopologue have been made available
through the \textsc{basecol}\footnote{\texttt{http://basecol.obspm.fr/}} database
\citep{Dubernet13}.
They are derived from a Maxwellian average over the corresponding hyperfine cross-sections
$\sigma_{\alpha\beta}$ \citep{Lique15}
\begin{equation} \label{eq:Mxav}
  R_{\alpha\rightarrow\beta}(T) = \left(\frac{8}{\pi\mu k_B^3 T^3}\right)^{1/2}
                                  \int_{0}^{\infty} \sigma_{\alpha\beta} E_c \ee^{-E_c/k_BT} \dd E_c \,,
\end{equation}
where $\mu$ is the reduced mass of the collision system, $E_c$ is the collision energy, and
$\alpha$, $\beta$ are the initial and final levels, respectively, each labelled with the $j,F_1,F$
quantum numbers.
These are obtained by coupling the rotational angular momentum with the two \isot{14}{N} nuclear
spins, e.g., $\mathbf{F}_1 = \mathbf{j} + \mathbf{I}_1$, and $\mathbf{F} = \mathbf{F}_1 + \mathbf{I}_2$.
The cross sections are, in turn, obtained by the recoupling technique starting from the ``spinless''
opacity tensor elements $P^K(j\rightarrow j')$:
\begin{equation} \label{eq:xsect}
  \sigma_{jF_1F\rightarrow j'F'_1F'} = \frac{\pi}{k_j^2}[F_1\,F'_1\,F'] 
           \sum_K \wsxj{F_1}{F'_1}{K}{F'}{F}{I_2}^2 \wsxj{j}{j'}{K}{F'_1}{F_1}{I_1}^2
          P^K(j\rightarrow j') \,.
\end{equation}
Here, $k_j$ is the wave-vector for the energy channel $E_c$, \mbox{$k_j^2 = (2\mu/\hbar^2)(E_c - E_j)$};
the terms in brace parentheses are the Wigner-$6j$ symbols, and the notation $[xy\ldots]$ is a handy
shorthand for the product $(2x + 1)(2y + 1)\ldots$\:\:. \par
Hyperfine cross-sections for the \nqn, and \qnn isotopic variant (one \isot{14}{N} nucleus)
are not included in the \textsc{basecol} compilation.
They can however be obtained, to a very good approximation, by summing Eq.~\eqref{eq:xsect} over
the final $F'$ states.
Using the orthogonality property of the $6j$ symbols it holds:
\begin{equation} \label{eq:orthog}
  \sum_{F'} [F'] \wsxj{F_1}{F'_1}{K}{F'}{F}{I_2}^2 = [F_1]^{-1} \,,
\end{equation}
because the triads ($F'_1,F_1,K$) and ($I_2,F,F'$) satisfy the triangular condition by definition
\citep{Daniel04}.
Hence, one has the equality:
\begin{equation} \label{eq:sumsig}
  \sum_{F'} \sigma_{jF_1F\rightarrow j'F'_1F'} = 
            \frac{\pi}{k_j^2}[F'_1] \wsxj{j}{j'}{K}{F'_1}{F_1}{I_1}^2 P^K(j\rightarrow j') =
            \sigma_{jF_1\rightarrow j'F'_1} \,,
\end{equation}
which, inserted in the \eqref{eq:Mxav} yields, to a very good approximation:
\begin{equation} \label{eq:sumR}
  \sum_{F'} R_{jF_1F\rightarrow j'F'_1F'}(T) = R_{jF_1\rightarrow j'F'_1}(T) \,.
\end{equation}
The \isot{15}{N}-bearing isotopologues contain only one quadrupolar nucleus and appropriate
angular momentum addition scheme is $\mathbf{F} = \mathbf{j} + \mathbf{I}$.
Thus, in the right-hand term of \eqref{eq:sumR}, the quantum number $F_1$ can be replaced by the
new $F$ to give the hyperfine coefficients for the \nqn/$p$-H$_2$ and \qnn/$p$-H$_2$ collisions.
In this treatment, we neglected the isotopic dependence of the cross sections, which is 
expected to be negligible for the $^{14}$N$\leftrightarrow^{15}$N substitution 
\citep[see for example][]{Buffa09}.

\par
The excitation rates, which are also required in MOLLIE, are derived through the detailed
balance relations
\begin{equation} \label{eq:db1}
  R_{jF_1F\leftarrow j'F'_1F'}(T) = \frac{\left[ F \right]}{\left[F'\right]} R_{jF_1F\rightarrow j'F'_1F'}(T) \, \ee^{-\Delta E/k_B T} \,,
  \quad \text{for \nnh} \,,
\end{equation}
\begin{equation} \label{eq:db2}
  R_{jF\rightarrow j'F'}(T) = \frac{\left[ F \right]}{\left[F'\right]} R_{jF\rightarrow j'F'}(T) \, \ee^{-\Delta E/k_B T} \,,
  \quad \text{for \nqn and \qnn} \,,
\end{equation}
where $\Delta E$ represents the energy difference between the hyperfine levels $(jF_1F)$ and 
$(j'F'_1F')$ or $(jF)$ and $(j'F')$.

\section{$\chi^2$ analysis \label{chi2App}}
In this Appendix, we show the $\chi^2$ values used to determine the best-fit value for the abundance (and thus for the column density) of each molecule, together with their uncertainties, in L183, L694-2 and L1544. The $\chi^2$ is evaluated according to Eq. \eqref{chi2}. L429 is not present due to the difficulties of its modeling. See the main text for more details.

\begin{figure}[h]
\centering
\includegraphics[width =0.5\textwidth]{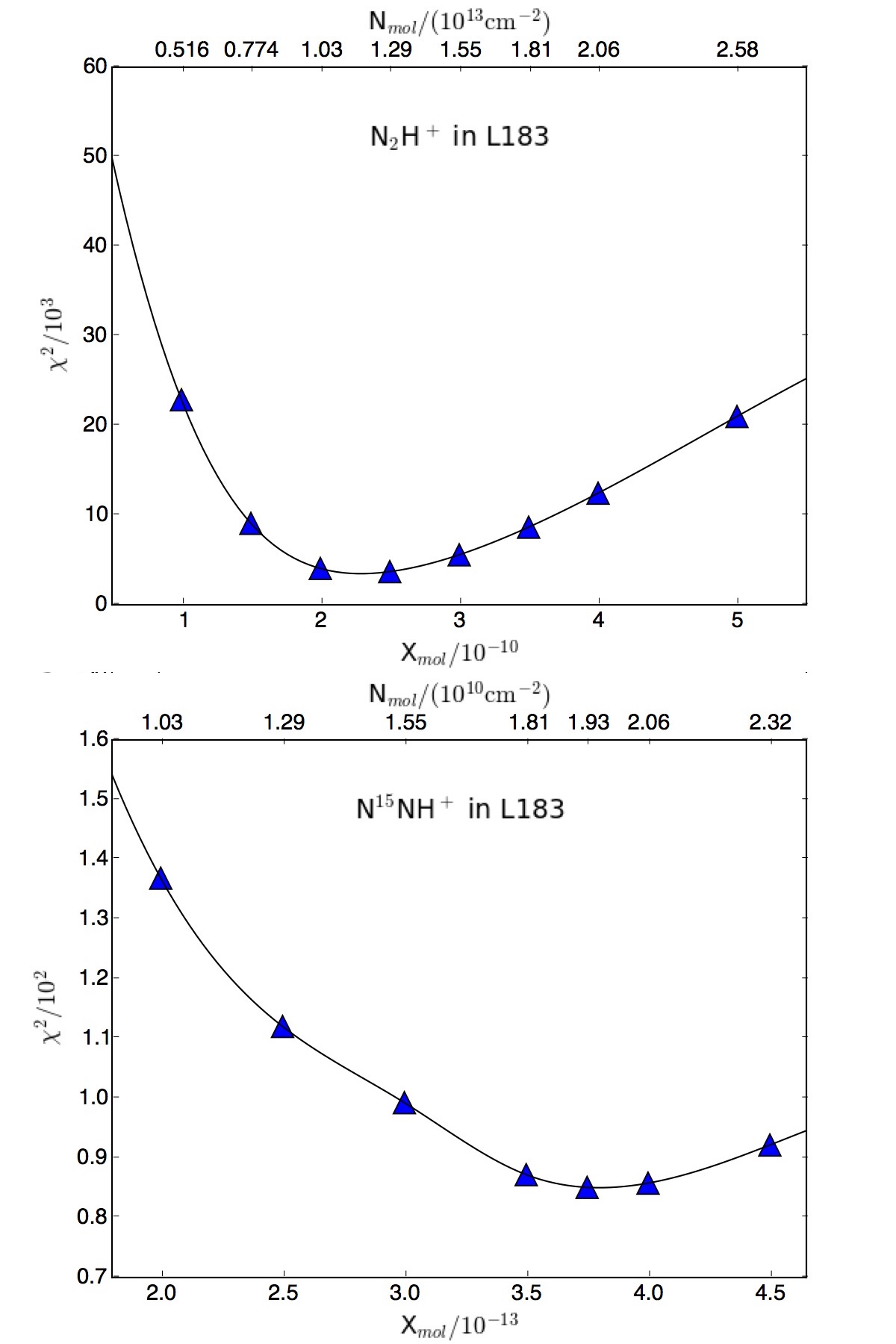}
\caption{The $\chi^2$ values for different abundance values (and corresponding column densities) in L183, for \nnh (upper panel) and \nqn (lower panel). The black curve is the one used to estimate the uncertainties, according to what said in Sec. \ref{MOLLIE}.}
\end{figure}

\begin{figure}[h]
\centering
\includegraphics[width =0.5\textwidth]{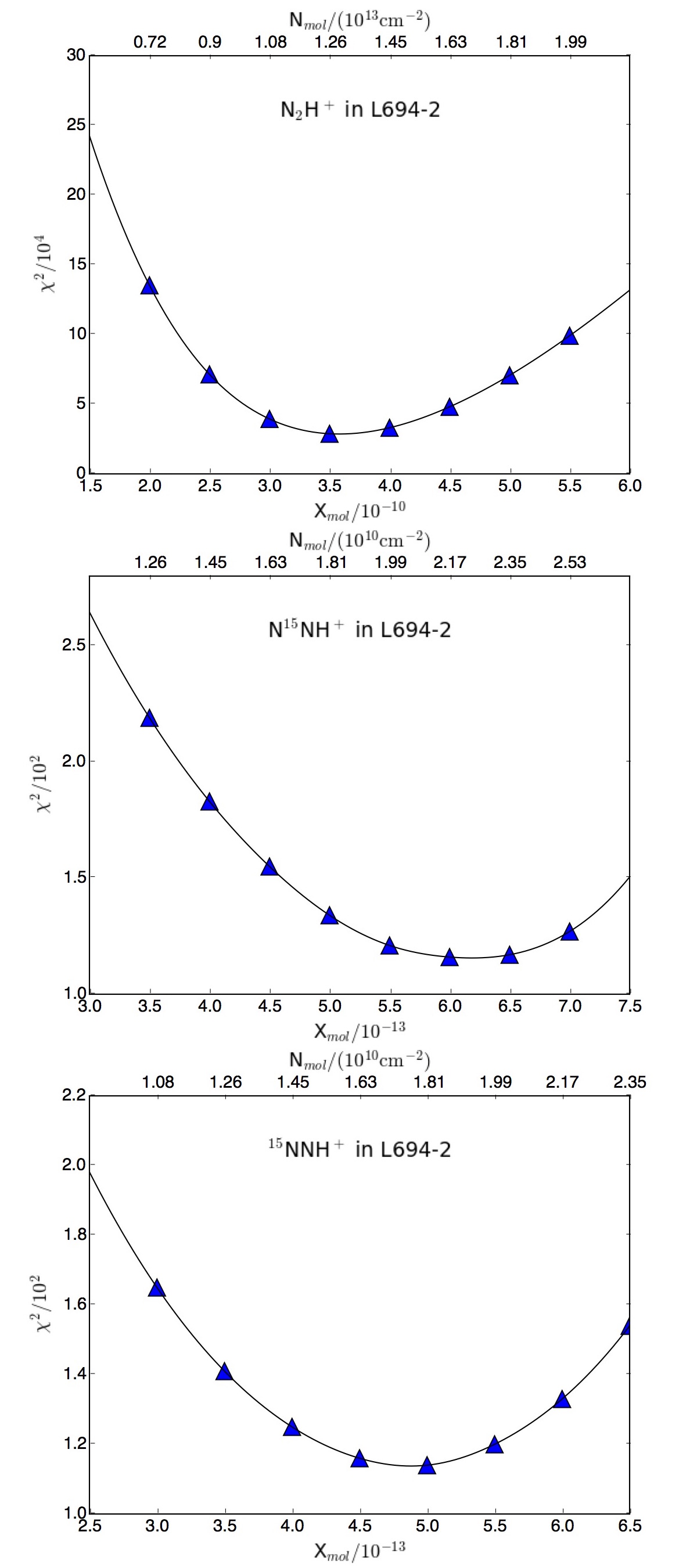}
\caption{The $\chi^2$ values for different abundance values (and corresponding column densities) for L694-2. From top to bottom: N$_2$H$^+$, N$^{15}$NH$^+$, $^{15}$NNH$^+$. The black curve is the one used to estimate the uncertainties, as explained in Sec. \ref{MOLLIE}.}
\end{figure}

\begin{figure}[h]
\centering
\includegraphics[width =0.5\textwidth]{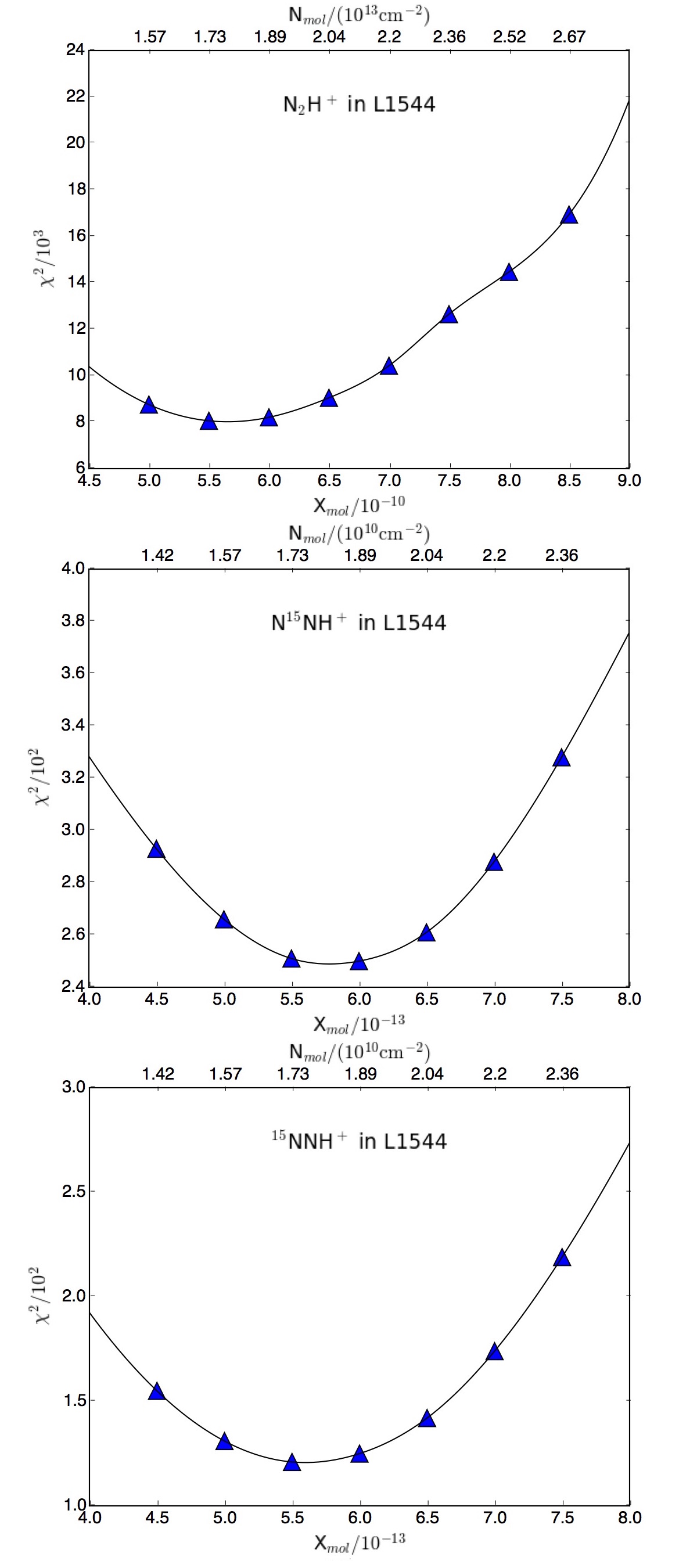}
\caption{The $\chi^2$ values obtained for different abundance values (and corresponding column densities) in L1544. From top to bottom: N$_2$H$^+$, N$^{15}$NH$^+$, $^{15}$NNH$^+$. The black curve is the one used to estimate the uncertainties, as explained in Sec. \ref{MOLLIE}.}
\end{figure}

\end{document}